\documentclass[twocolumn,showpacs,preprintnumbers,amsmath,amssymb]{revtex4}

\usepackage{graphicx}
\usepackage{dcolumn}
\usepackage{bm}
\usepackage{color}
\newcommand{\ua}{\uparrow}
\newcommand{\da}{\downarrow}
\newcommand{\dg}{\dagger}

\begin{document}

\preprint{}

\title{Monte Carlo study of an unconventional superconducting phase in Ir-oxide $J_{\rm eff}=1/2$ Mott insulators induced by carrier doping}

\author{Hiroshi Watanabe}
 \email{h-watanabe@riken.jp}
\author{Tomonori Shirakawa}
\author{Seiji Yunoki}
\affiliation{%
Computational Condensed Matter Physics Laboratory, RIKEN ASI, Wako, Saitama 351-0198, Japan\\
CREST, Japan Science and Technology Agency, Kawaguchi, Saitama 332-0012, Japan\\
Computational Materials Science Research Team, RIKEN AICS, Kobe, Hyogo 650-0047, Japan
}%

\date{\today}

\begin{abstract}
Based on a microscopic theoretical study, we show that novel superconductivity is induced by carrier doping in layered 
perovskite Ir oxides where a strong spin-orbit 
coupling causes an effective total angular momentum $J_{\rm eff}=1/2$ Mott insulator. Using a variational Monte Carlo method, 
we find an unconventional superconducting state in the ground state phase diagram of a $t_{2g}$ three-orbital Hubbard model on 
the square lattice. 
This superconducting state is characterized by a $d_{x^2-y^2}$-wave ``pseudospin singlet'' formed by the $J_{\rm eff}=1/2$ Kramers doublet, 
which thus contains inter-orbital as well as both singlet and triplet components of $t_{2g}$ electrons.  
The superconducting state is found stable only by electron doping, but not by hole doping, for the case of carrier doped Sr$_2$IrO$_4$.  
We also study an effective single-orbital Hubbard model to discuss the similarities to high-$T_{\rm c}$ cuprate superconductors 
and the multi-orbital effects. 

\end{abstract}

\maketitle

Search for novel superconductivity (SC) is one of the most interesting and fundamental issues in condensed matter physics.
In strongly correlated electron systems, SC is very often induced in the vicinity of long-range ordered states, 
thus suggesting the importance of the enhanced fluctuations for the SC. 
The most studied example is found in high-$T_{\mathrm{c}}$ cuprate superconductors~\cite{Bednorz} where the SC occurs next to an antiferromagnetic (AF) Mott insulator, 
and thus the AF spin fluctuations are often believed to be responsible for the SC~\cite{cuprate}. 

Recently, layered perovskite $5d$ transition metal oxides Sr$_2$IrO$_4$~\cite{Randall,Cao} and Ba$_2$IrO$_4$~\cite{Okabe1} 
have attracted much attention because several experiments have revealed a novel spin-orbit-induced $J_{\mathrm{eff}}=1/2$ Mott 
insulating behavior at low temperatures~\cite{Kim1,Kim2,Ishii,Okabe2,Kim3,Cetin,Fujiyama}. 
In these systems, due to a large spin-orbit coupling (SOC) and a large crystal field splitting, 
the local electronic state with nominally $(t_{2g})^5$ electron configuration in Ir ion is represented 
by an effective total angular momentum $J_{\mathrm{eff}}$=$\left|-\bm{L}+\bm{S}\right|$=1/2~\cite{sugano}. 
In these insulators with effectively one hole per Ir ion, this pseudospin remains a good quantum number and orders antiferromagnetically. 
Indeed, very recent experiments in Sr$_2$IrO$_4$ have observed that 
the low-energy magnetic excitations can be well described by a pseudospin 1/2 AF Heisenberg model 
with AF exchange coupling $J_{\mathrm{ex}}$=60-100 meV~\cite{Kim3,Fujiyama}. 
The theoretical studies also support this $J_{\mathrm{eff}}=1/2$ Mott insulator in these Ir oxides~\cite{Jackeli,Jin,Watanabe,Martins,Shirakawa}. 

It is now interesting to compare Sr$_2$IrO$_4$ with parent compunds of high-$T_{\mathrm{c}}$ cuprate superconductors such as La$_2$CuO$_4$. 
The similarities are summarized as follows: 
(i) both are in the same layered perovskite structure of K$_2$NiF$_4$ type, i.e., in a quasi two-dimensional (2D) structure, 
(ii) both have effectively one hole per Ir or Cu ion, 
(iii) both show spin or pseudospin 1/2 AF order at low temperatures with low-energy magnetic excitations described 
by a spin or pseudospin 1/2 AF Heisenberg model, 
and (iv) both have the large $J_{\mathrm{ex}}$ of the same order. Considering these similarities, 
it is tempting to expect that novel unconventional SC with possibly a high critical temperature ($T_{\rm c}$) is induced once mobile 
carriers are introduced into the $J_{\mathrm{eff}}=1/2$ Mott insulating  Sr$_2$IrO$_4$.
Although there have been several reports suggesting this possibility~\cite{Kim3,Wang}, 
it is highly desirable to show, based on microscopic calculations, the existence of SC in the doped $J_{\mathrm{eff}}=1/2$ Mott insulator. 

In this Letter, using a variational Monte Carlo (VMC) method, we study the ground state phase diagram of a $t_{2g}$ three-orbital 
Hubbard model, and show that novel unconventional SC is induced by carrier doping in the $J_{\mathrm{eff}}=1/2$ Mott insulator. 
This SC is characterized by a $d_{x^2-y^2}$-wave ``pseudospin singlet" formed by the $J_{\mathrm{eff}}=1/2$ Kramers doublet, 
which thus consists of inter-orbital pairings and both singlet and triplet pairings of $t_{2g}$ electrons. 
We also find that the SC is stable only by electron doping, but not by hole doping, for the case of carrier doped Sr$_2$IrO$_4$. 
We furthermore study an effective single-orbital Hubbard model to discuss the similarities to high-$T_{\rm c}$ cuprates 
and the multi-orbital effects. 

One of the simplest models for the Ir oxides, which we shall study here, is 
a $t_{2g}$ three-orbital Hubbard model on the square lattice defined by 
$H=H_{\mathrm{kin}}+H_{\mathrm{SO}}+H_{\mathrm{I}}$, where 
$H_{\mathrm{kin}}=\sum_{\bm{k},\alpha,\sigma}\varepsilon_{\alpha}(\bm{k})c_{\bm{k}\alpha\sigma}^{\dagger}
        c_{\bm{k}\alpha\sigma}$ is the kinetic term, 
$H_{\mathrm{SO}}=\lambda\sum_{i}\bm{L}_i\cdot\bm{S}_i$ is the SOC term with a coupling constant $\lambda$, and 
$
H_{\mathrm{I}}=U\sum_{i,\alpha}n_{i\alpha\uparrow}n_{i\alpha\downarrow}
     +\sum_{i,\alpha<\beta,\sigma}\left[U'n_{i\alpha\sigma}n_{i\beta\bar{\sigma}}
  +(U'-J)n_{i\alpha\sigma}n_{i\beta\sigma}\right]
 +J\sum_{i,\alpha<\beta}(c^{\dagger}_{i\alpha\uparrow}c^{\dagger}_{i\beta\downarrow}
  c_{i\alpha\downarrow}c_{i\beta\uparrow}
  +c^{\dagger}_{i\alpha\uparrow}c^{\dagger}_{i\alpha\downarrow}
   c_{i\beta\downarrow}c_{i\beta\uparrow}+\mathrm{H.c.})
   \label{int}
$
is the Coulomb interaction term including intra-orbital ($U$), inter-orbital ($U'$), and spin-flip and pair-hopping ($J$) interactions~\cite{Watanabe}. 
Here, $c^\dag_{i\alpha\sigma}$ is an electron creation operator at site $i$ with spin $\sigma(=\uparrow,\downarrow)$ 
and orbital $\alpha(=yz,zx,xy)$, $n_{i\alpha\sigma}=c^\dag_{i\alpha\sigma}c_{i\alpha\sigma}$, and $c^\dag_{{\bm k}\alpha\sigma}$ 
is the Fourier transformation of $c^\dag_{i\alpha\sigma}$. We impose $U=U'+2J$ for rotational symmetry~\cite{Kanamori}.  

The kinetic and the SOC terms can be combined, $H_0(t_i,\mu_{xy}, \lambda)=H_{\mathrm{kin}}+H_{\mathrm{SO}}$, in the matrix form 
\begin{align}
H_0&=\sum_{\bm{k},\sigma}\left(c^{\dagger}_{\bm{k}yz\sigma},c^{\dagger}_{\bm{k}zx\sigma},c^{\dagger}_{\bm{k}xy\bar{\sigma}}\right) \notag \\
&\;\;\;\;\;\times
\begin{pmatrix}
\varepsilon_{yz}(\bm{k}) & \mathrm{i}s_\sigma\lambda/2 & -s_\sigma\lambda/2 \\
-\mathrm{i}s_\sigma\lambda/2 & \varepsilon_{zx}(\bm{k}) & \mathrm{i}\lambda/2 \\
-s_\sigma\lambda/2 & -\mathrm{i}\lambda/2 & \varepsilon_{xy}(\bm{k})
\end{pmatrix}
\begin{pmatrix}
c_{\bm{k}yz\sigma} \\
c_{\bm{k}zx\sigma} \\
c_{\bm{k}xy\bar{\sigma}}
\end{pmatrix} \\ \notag
&=\sum_{\bm{k},m, s}E_m(\bm{k})a^{\dg}_{\bm{k}m s}
a_{\bm{k}m s},  
\end{align}
where $\bar\sigma$ is the opposite spin of $\sigma$ and $s_\sigma=1(-1)$ for $\sigma=\uparrow(\downarrow)$.  
Notice that the SOC mixes the different electron spins ($\sigma$ and $\bar{\sigma}$), and the new quasiparticles, 
obtained by diagonalizing $H_0$, are characterized by band index $m(=1,2,3)$ 
and pseudospin $s=(\uparrow,\downarrow)$ with a creation operator $a^{\dagger}_{\bm{k}m s}$. 
In the atomic limit with $\varepsilon_{yz}(\bm{k})=\varepsilon_{zx}(\bm{k})=\varepsilon_{xy}(\bm{k})=0$, 
the sixfold degenerate $t_{2g}$ levels are split into twofold degenerate $J_{\mathrm{eff}}=1/2$ states ($m=1$) 
and fourfold degenerate $J_{\mathrm{eff}}=3/2$ states ($m=2,3$)~\cite{sugano}. 
The undoped filling corresponds to electron density $n=5$, 
and in the atomic limit all states but the $J_{\mathrm{eff}}=1/2$ states are fully occupied. 

\begin{figure}[t]
\begin{center}
\includegraphics[width=0.85\hsize]{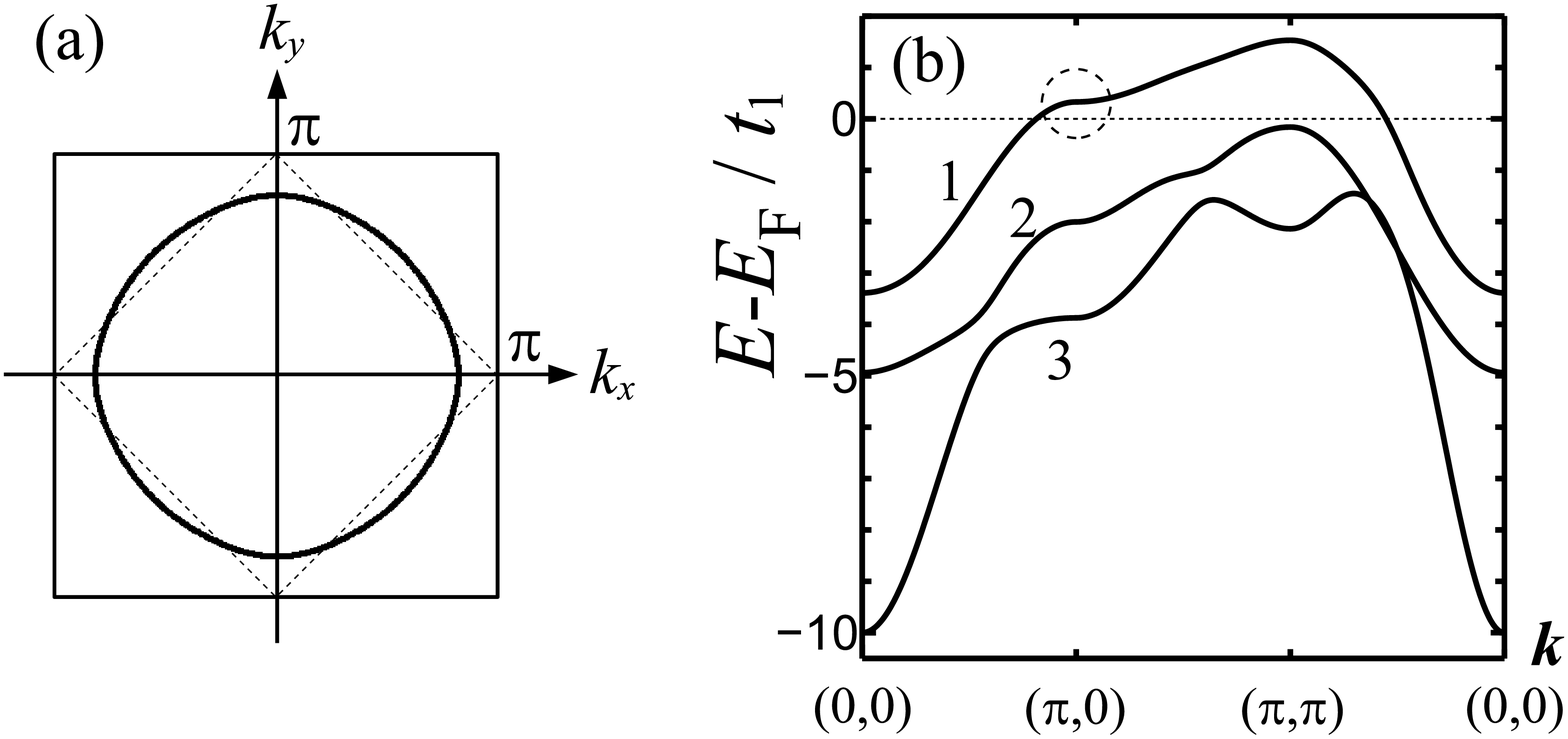}
\caption{\label{fig1} 
(a) Fermi surface and (b) energy dispersions of the non-interacting tight-binding energy band for Sr$_2$IrO$_4$ with electron density $n=5$. 
Numbers in (b) denote the band index $m$, and $E_{\rm F}$ is the Fermi energy. 
A set of tight-binding parameters used is $(t_2, t_3, t_4, t_5, \mu_{xy}, \lambda) = (0.5,0.25,1.03,0.17,-1.0,1.39)t_1$. 
} 
\end{center}
\end{figure}

In Ref.~\onlinecite{Watanabe}, 
we have constructed the non-interacting tight-binding energy band for Sr$_2$IrO$_4$: 
$\varepsilon_{yz}(\bm{k})=-2t_5\cos k_x-2t_4\cos k_y$,
$\varepsilon_{zx}(\bm{k})=-2t_4\cos k_x-2t_5\cos k_y$, and
$\varepsilon_{xy}(\bm{k})=-2t_1(\cos k_x+\cos k_y)-4t_2\cos k_x\cos k_y-2t_3(\cos 2k_x+\cos 2k_y)+\mu_{xy}$  
with a set of tight-binding parameters 
($t_1, t_2, t_3, t_4, t_5, \mu_{xy}, \lambda$)=(0.36, 0.18, 0.09, 0.37 0.06, $-0.36$, 0.50) eV. 
The corresponding Fermi surface and energy dispersions are shown in Fig.~\ref{fig1}. 
As shown in Fig.~\ref{fig1} (b), 
we assign the band index $m=1$, 2, and 3 from the highest band to the lowest one, and 
only band 1 crosses the Fermi Energy~\cite{note1,note2}. 

The effect of Coulomb interactions is treated using a VMC method~\cite{Watanabe}. 
The trial wave function $\left|\Psi \right>$ considered here is composed of three parts: 
$\left|\Psi \right>=P_{\mathrm{J_c}}P^{(3)}_{\mathrm{G}}\left|\Phi \right>$.
$\left|\Phi \right>$ is the one-body part obtained by diagonalizing $\tilde H_0= H_0(\tilde{t}_i, \tilde{\mu}_{xy}, \tilde{\lambda}_{\alpha\beta})$ 
with variational ``renormalized" tight-binding parameters $\{ \tilde{t}_i, \tilde{\mu}_{xy}, \tilde{\lambda}_{\alpha\beta} \}$. 
Notice that we introduce an orbital dependent ``effective'' SOC constant: $\lambda\rightarrow\tilde{\lambda}_{\alpha\beta}$.
To treat magnetically ordered states, a term with a different magnetic order parameter is added to $\tilde H_0$.
Here, we consider out-of-plane AF order (along $z$ axis, $z$-AF) and in-plane AF order (along $x$ axis, $x$-AF), described respectively by 
$\sum_{i,m} M^z_m\mathrm{e}^{\mathrm{i}\bm{Q}\cdot\bm{r}_i}(a^{\dg}_{im\ua}a_{im\ua}
-a^{\dg}_{im\da}a_{im\da})$ and 
$\sum_{i,m} M^x_m\mathrm{e}^{\mathrm{i}\bm{Q}\cdot\bm{r}_i}(a^{\dg}_{im\ua}a_{im\da}
+a^{\dg}_{im\da}a_{im\ua})$, where 
$a^{\dagger}_{ims}$ is the Fourier transformation of $a^{\dagger}_{\bm{k}ms}$ and $\bm{Q}=(\pi,\pi)$. 
The order parameters ($M^z_1,M^z_2,M^z_3$) for $z$-AF and ($M^x_1,M^x_2,M^x_3$) for $x$-AF are variational parameters.
With an appropriate basis transformation, we obtain the original $t_{2g}$ orbital representation in real space 
and construct the Slater determinant $\left|\Phi \right>$ for VMC simulation.

To study a possible superconducting state, we consider the following BCS-type Hamiltonian,

\begin{align}
\tilde H_{\mathrm{BCS}}&=\sum_{\bm{k}}
    \left( a_{\bm{k}1\ua}^{\dg}, a_{\bm{k}2\ua}^{\dg}, a_{\bm{k}3\ua}^{\dg},  
   a_{-\bm{k}1\da}, a_{-\bm{k}2\da}, a_{-\bm{k}3\da}\right) \times \notag \\
  &\begin{pmatrix}
   \xi_1 & 0 & 0 & \varDelta_{11} & \varDelta_{12} & \varDelta_{13} \\ 
   0 & \xi_2 & 0 & \varDelta_{21} & \varDelta_{22} & \varDelta_{23} \\
   0 & 0 & \xi_3 & \varDelta_{31} & \varDelta_{32} & \varDelta_{33} \\
   \varDelta^*_{11} & \varDelta^*_{12} & \varDelta^*_{13} & -\xi_1 & 0 & 0 \\
   \varDelta^*_{21} & \varDelta^*_{22} & \varDelta^*_{23} & 0 & -\xi_2 & 0 \\
   \varDelta^*_{31} & \varDelta^*_{32} & \varDelta^*_{33} & 0 & 0 & -\xi_3 
 \end{pmatrix}
 \begin{pmatrix}
a_{\bm{k}1\ua} \\ a_{\bm{k}2\ua} \\ a_{\bm{k}3\ua} \\  
   a_{-\bm{k}1\da}^{\dg}\\ a_{-\bm{k}2\da}^{\dg}\\ a_{-\bm{k}3\da}^{\dg} \end{pmatrix},  \label{BCS}
\end{align} 
where 
$\xi_m=\tilde{E}_m(\bm{k})-\tilde{\mu}$, $\tilde{E}_m(\bm{k})$ is the eigenvalues of $\tilde H_0$, 
and $\tilde{\mu}$ is a variational parameter of the chemical potential form. 
The gap functions $\varDelta_{mm'}$ (${\bm k}$ dependence implicitly assumed) are additional variational parameters. 
After diagonalizing $\tilde H_{\mathrm{BCS}}$, the ground state of the superconducting state is obtained 
by creating all negative energy states ($\bar{\gamma}^{\dagger}_{\nu}$) 
and annihilating all positive energy states ($\gamma_{\nu}$) on the vacuum state, 
$\left|\Phi\right>=\prod_{\nu}\gamma_{\nu}\bar{\gamma}^{\dagger}_{\nu}\left|0\right>$. 
In this study, we consider mostly the intra-band (but inter-orbital) pairing, 
namely, $\varDelta_{mm'}=0$ for $m\neq m'$~\cite{note3}.

The operator $P^{(3)}_{\mathrm{G}}$ is a Gutzwiller factor extended for the three-orbital system 
and the operator $P_{\mathrm{J_c}}$ is a long-rang charge Jastrow factor. These operators 
are exactly the same ones reported in Ref.~[\onlinecite{Watanabe}]. 
The ground state energies are calculated with a VMC method.
The variational parameters, as many as 80 parameters for a 20$\times$20 square lattice, are 
simultaneously optimized to minimize the variational energy 
by using the stochastic reconfiguration method~\cite{Sorella}. 


\begin{figure}
\includegraphics[width=0.75\hsize]{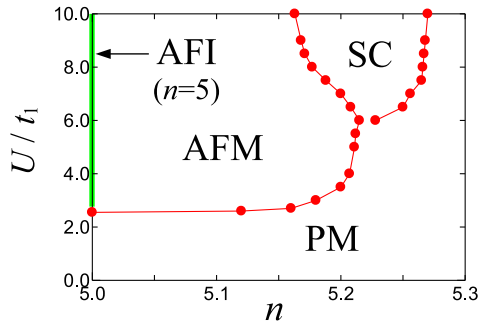}
\caption{\label{fig2} (color online) The ground state phase diagram of the three-orbital Hubbard model on the 2D 20$\times$20 square lattice with $J=0$. 
PM, AFI, AFM, and SC denote paramagnetic metal, AF insulator, AF metal, and $d_{x^2-y^2}$-wave ``pseudospin singlet" SC, 
respectively.}
\end{figure}


Let us first summarize in Fig.~\ref{fig2} our main results for the ground state phase diagram where the electron density $n$ and 
the intra-orbital Coulomb interaction $U$ are varied. In this phase diagram, we set the Hund's coupling $J=0$. 
Therefore, the pseudospin rotational symmetry is 
preserved~\cite{Jackeli,Watanabe}, and $z$-AF and $x$-AF are energetically degenerate (denoted simply by AF in the phase diagram). 
At $n=5$, the AF insulator appears for $U/t_1\agt2.6$. This insulating state is considered to be the spin-orbit-induced $J_{\rm eff}=1/2$ 
Mott insulator observed in Sr$_2$IrO$_4$~\cite{Watanabe}. 
With increasing $n$ by electron doping, the AF order is eventually destroyed, and it is replaced by the superconducting state for $U/t_1\agt6$ 
and $n\sim5.2$ (namely, $\sim20$\% electron doping). 
The Cooper pair symmetry of this SC is found to be a $d_{x^2-y^2}$-wave ``pseudospin singlet'' formed by the 
$J_{\rm eff}=1/2$ Kramers doublet~\cite{note4}. 
Thus, this pairing contains inter-orbital components as well as both singlet and triplet components of $t_{2g}$ electrons. 
This can be easily seen in the limit of large SOC ($\lambda\rightarrow\infty$), where the most dominant Cooper pair is expressed by 
$
a_{\bm{k}1\ua}^{\dagger}a_{-\bm{k}1\da}^{\dagger} \propto 
\left( c_{\bm{k}xy\ua}^{\dagger} + c_{\bm{k}yz\da}^{\dagger} + \mathrm{i}c_{\bm{k}zx\da}^{\dagger} \right)
\left( c_{-\bm{k}xy\da}^{\dagger} - c_{-\bm{k}yz\ua}^{\dagger} + \mathrm{i} c_{-\bm{k}zx\ua}^{\dagger} \right). 
$
When $\lambda$ is finite, the coefficient of each term is different from the one in the above limit and it is determined 
variationally. 
It should be also noted that the superconducting order parameters considered in Eq.~(\ref{BCS}) include not only the nearest-neighbor pairing but also 
long-range pairings, e.g.,  up to the 5th neighbor 
for the $d_{x^2-y^2}$-wave symmetry~\cite{note5}. 
It is known that the simplest $d_{x^2-y^2}$-wave pairing ($\propto\cos k_x-\cos k_y$) is significantly modified by the long-range 
contribution in the underdoped regime of high-$T_{\mathrm{c}}$ cuprates~\cite{Mesot}.
The long-range pairings are important also in our three-orbital model and give the lower variational energy for the superconducting state.

\begin{figure}[t]
\includegraphics[width=0.75\hsize]{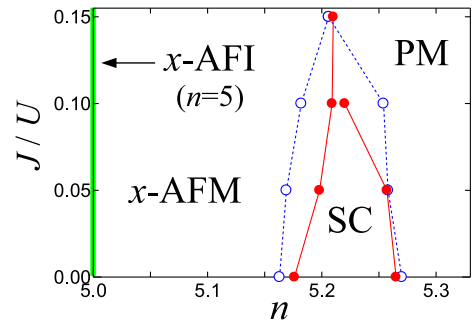}
\caption{\label{fig3}
(color online) The ground state phase diagram of the three-orbital Hubbard model on the 2D 20$\times$20 square lattice. 
The solid-red and dotted-blue lines represent the phase boundaries for $U/t_1=8$ and 10, respectively.
PM, $x$-AFI, $x$-AFM, and SC denote paramagnetic metal, in-plane AF insulator, in-plane AF metal, and $d_{x^2-y^2}$-wave ``pseudospin singlet" 
SC, respectively. }
\end{figure}

To the contrary, with decreasing electron density by hole doping for $n<5$, we do not find a superconducting state in the phase 
diagram (not shown here), where instead a AF metal dominates the SC~\cite{note6}. 
This electron-hole asymmetry of the phase diagram reminds us of the phase diagram of a model for high-$T_{\mathrm{c}}$ cuprates. 
As in the case of high-$T_{\mathrm{c}}$ cuprates, the asymmetry found here for the three-orbital Hubbard model is understood due to 
a band structure effect. 
As seen in Fig.~\ref{fig1} (b) (indicated by the dotted circle), the dispersion of band $m=1$ is flat for ${\bm k}$ around $(\pm\pi,0)$ and 
(0,$\pm\pi$), which induces van Hove singularity in the density of states (DOS) for $n>5$ (but not for $n<5$). 
The large DOS originated from this flat dispersion favors the SC~\cite{Dagotto}. 

Next, we study the effect of Hund's coupling $J$. Fig.~\ref{fig3} shows the ground state phase diagram in a $n$-$J/U$ plane for $U/t_1=8$. 
The introduction of a finite $J$ breaks the pseudospin rotational sysmetry~\cite{Watanabe,Jackeli}, and the states with 
in-plane AF order are favored over those with out-of-plane AF order (see Fig.~\ref{fig3}). 
We also find in Fig.~\ref{fig3} that the superconducting state remains stable, but the superconducting region gradually reduces and eventually disappears with 
further increasing $J/U$. 
This implies that the Hund's coupling $J$ unfavors the SC. 
To understand the effect of Hund's coupling $J$, we recall that the charge gap $\Delta_{\mathrm{c}}$ in the limit of strong Coulomb 
interactions with $d^5$ configuration is 
$\Delta_{\mathrm{c}}=E(d^4)+E(d^6)-2E(d^5)=U-3J$~\cite{deMedici}. 
Thus, the effect of the Hund's coupling $J$ is to reduce the effective electron correlations. 
Comparing the variational energies of the paramagnetic metal and the superconducting state, 
we find that the SC is stabilized by the gain of the interaction energies at the expense of the band energies. 
Therefore, with increasing the Hund's coupling $J$, the condensation energy of the SC is greatly reduced and the SC is eventually destabilized~\cite{tba}.   
We also study the $J/U$ dependence of the SC for $U/t_1=10$, and the results are shown in Fig.~\ref{fig3}. 
Although the Hund's coupling $J$ is still destructive for the SC, the superconducting region is found to be rather extended as compared with the results 
for $U/t_1=8$. 
This suggests that the SC is more likely to be found experimentally in the doped Ir oxides with larger $U/t_1$. 

It should be noted here that the realistic values of the Coulomb interactions for the Ir oxides are still 
controversial, and that even these values may vary greatly for different models because the screening mechanisms 
can be different~\cite{Vaugier}. Nevertherless, the previous studies~\cite{Martins,Arita,Comin} have reported that 
$U=2$--$3$ eV (i.e., $U/t_1=5.6$--$8.3$) and $J/U=0.05$--$0.20$, which are 
still within (or at least in the vicinity of) the parameter region where we find the SC induced by electron doping. 

Finally, let us study an effective single-orbital Hubbard model~\cite{Jin,Wang} to discuss the similarities to high-$T_{\mathrm{c}}$ cuprates 
and the multi-orbital effects. 
The effective single-orbital model is readily constructed by fitting band $m=1$, i.e., ``$J_{\mathrm{eff}}=1/2$ band", in Fig.~\ref{fig1} (b)
using the dispersion relation $\varepsilon(\bm{k})=-2t(\cos k_x+\cos k_y)-4t'\cos k_x\cos k_y-2t''(\cos 2k_x+\cos 2k_y)$ 
with $(t, t', t'')=(0.221, 0.057, -0.011)$ eV. 
The value of $U/t$ for the effective single-orbital Hubbard model is chosen to be 13 ($\approx8t_1/t$), 
which should correspond to $U/t_1=8$ for the three-orbital Hubbard model. 

The ground state phase diagram of this effective single-orbital Hubbard model is studied using the VMC method. 
The trial wave function used is almost the same as the one for the three-orbital model: 
$\left|\Psi \right>=P_{\mathrm{J_c}}P^{(1)}_{\mathrm{G}}\left|\Phi \right>$. 
The operator $P^{(1)}_{\mathrm{G}}=\prod_{i}\left[1-(1-g)n_{i\ua}n_{i\da}\right]$ is a usual Gutzwiller factor with only one variational 
parameter $g$. For the one-body part $\left|\Phi \right>$, we consider the BCS-type wave function with both AF and $d_{x^2-y^2}$-wave 
superconducting orders~\cite{Giamarchi}. 

\begin{figure}
\includegraphics[width=0.75\hsize]{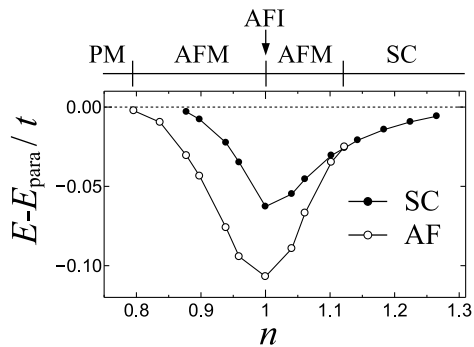}
\caption{\label{fig4} 
Energy comparison of the effective single-orbital Hubbard model on the 2D 14$\times$14 square lattice with $U/t=13$, $t'/t=0.26$, and $t''/t=-0.05$. 
The energies for different states are measured from the one for the paramagnetic state. The phase diagram is indicated 
in the upper part of the figure.  Here $n=1$ corresponds to half filling. PM, AFI, AFM, and SC stand for paramagnetic metal, AF insulator, 
AF metal, and $d_{x^2-y^2}$-wave SC, respectively. }
\end{figure}

Fig.~\ref{fig4} shows the doping dependence of the variational energies for different states in the single-orbital Hubbard model. 
At half filling ($n=1$), the ground state is an AF insulator. 
By electron doping with $n>1$, the AF order vanishes around 12$\%$ doping and the $d_{x^2-y^2}$-wave SC appears. 
To the contrary, the SC is absent in the hole-doped side ($n<1$).
This phase diagram is very similar to the one of the three-orbital Hubbard model and also to the one of a model for 
high-$T_{\rm c}$ cuprates~\cite{Kobayashi}, suggesting that the mechanisms of SC are the same origin for both systems. 
It should be noted that the electron-hole asymmetry found in a model for high-$T_{\mathrm{c}}$ cuprates, 
where the hole doping favors the SC more than the electron doping, is opposite to the results obtained here (see Fig.~\ref{fig4}). 
This difference is simply because of the sign difference of $t'/t$ between the effective single-orbital model ($t'/t>0$) and a model 
for high-$T_{\mathrm{c}}$ cuprates ($t'/t<0$)~\cite{Eskes}. 

Although they are qualitatively similar, the phase diagrams for the three-orbital and the effective single-orbital Hubbard models are 
quantitatively different. For instance, the region of the SC in the single-orbital model seems larger than that in the three-orbital model
(Fig.~\ref{fig4} should be compared with Fig.~\ref{fig2} at $U/t_1=8$).
This indicates that a multi-orbital effect is to destruct the SC. 
The reason can be attributed to the reduction of the effective electron correlations due to large orbital fluctuations, which 
certainly decrease the probability of facing double occupancy at the same orbital. 
As mentioned above, the reduction of the effective electron correlations destabilizes the SC and its region becomes smaller. 

In summary, we have studied the ground state phase diagram of the three-orbital Hubbard model for Sr$_2$IrO$_4$. 
We have found the unconventional SC induced by carrier doping in the $J_{\rm eff}=1/2$ Mott insulator. This SC is 
characterized by the $d_{x^2-y^2}$-wave ``pseudospin singlet" formed by the $J_{\rm eff}=1/2$ Kramers doublet. 
We have shown that the SC is induced only by electron doping, but not by hole doping, for the case of carrier doped Sr$_2$IrO$_4$. 
By studying the effective single-orbital Hubbard model constructed from the ``$J_{\rm eff}=1/2$ band", we have found the similar phase diagram
to the one of a model for 
high-$T_{\mathrm{c}}$ cuprates, suggesting the same mechanism of the SC in both systems. Finally, it should be noted that SC has not 
been observed yet experimentally in layered perovskite Ir oxides. We hope that our study will stimulate further experimental as well as theoretical 
studies in this direction. 
 
The authors thank Y. Yanase, J. Akimitsu, M. Isobe, H. Okabe, and S. Fujiyama for useful discussions.
The computation has been done using the RIKEN Cluster of Clusters (RICC) facility and the facilities of the Supercomputer Center,
Institute for Solid State Physics, University of Tokyo.
This work has been supported by Grant-in-Aid for Scientific Research from MEXT Japan under the grant numbers 24740251 and 24740269.


\begin{thebibliography}{99} 
\bibitem{Bednorz} J. G. Bednorz and K. A. Muller, Z. Phys. B \textbf{64}, 189 (1986).
\bibitem{cuprate} D. J. Scalapino, Phys. Rept. \textbf{250}, 329 (1995);
T. Moriya and K. Ueda, Adv. Phys. \textbf{49}, 555 (2000).
\bibitem{Randall} J. J. Randall \textit{et al}., J. Am. Chem. Soc. \textbf{79}, 266 (1957).
\bibitem{Cao} 
R. J. Cava, B. Batlogg, K. Kiyono, H. Takagi, J. J. Krajewski, W. F. Peck, Jr., L. W. Rupp, Jr., and C. H. Chen, 
Phys. Rev. B {\bf 49}, 11890 (1994); 
T. Shimura, Y. Inaguma, T. Nakamura, M. Itoh, and Y. Morii, Phys. Rev. B {\bf 52}, 9143 (1995);  
G. Cao, J. Bolivar, S. McCall, J. E. Crow, and R. P. Guertin, Phys. Rev. B {\bf 57}, R11039 (1998). 
\bibitem{Okabe1} H. Okabe, M. Isobe, E. Takayama-Muromachi, A. Koda, S. Takeshita, M. Hiraishi, M. Miyazaki, R. Kadono, 
Y. Miyake, and J. Akimitsu, Phys. Rev. B \textbf{83}, 155118 (2011).
\bibitem{Kim1} B. J. Kim, Hosub Jin, S. J. Moon, J.-Y. Kim, B.-G. Park, C. S. Leem, J. Yu, T. W. Noh, C. Kim, S.-J. Oh,
J.-H. Park, V. Durairaj, G. Cao, and E. Rotenberg, Phys. Rev. Lett. \textbf{101}, 076402 (2008).
\bibitem{Kim2} B. J. Kim, H. Ohsumi, T. Komesu, S. Sakai, T. Morita, H. Takagi, T. Arima, 
Science \textbf{323}, 1329 (2009).
\bibitem{Ishii} K. Ishii, I. Jarrige, M. Yoshida, K. Ikeuchi, J. Mizuki, K. Ohashi, T. Takayama, J. Matsuno, and H. Takagi, 
Phys. Rev. B {\bf 83}, 115121 (2011).  
\bibitem{Okabe2} H. Okabe, N. Takeshita, M. Isobe, E. Takayama-Muromachi, T. Muranaka, and J. Akimitsu, 
Phys. Rev. B \textbf{84}, 115127 (2011).
\bibitem{Kim3} J. Kim, D. Casa, M. H. Upton, T. Gog, Y.-J. Kim, J. F. Mitchell, M. van Veenendaal, M. Daghofer, J. van den Brink, 
G. Khaliullin, and B. J. Kim, Phys. Rev. Lett. {\bf 108}, 177003 (2012). 
\bibitem{Cetin} M. F. Cetin, P. Lemmens, V. Gnezdilov, D. Wulferding, D. Menzel, T. Takayama, K. Ohashi, and H. Takagi, 
Phys. Rev. B \textbf{85}, 195148 (2012). 
\bibitem{Fujiyama} S. Fujiyama, H. Ohsumi, T. Komesu, J. Matsuno, B.J. Kim, M. Takata, T. Arima, and H. Takagi, 
Phys. Rev. Lett. \textbf{108}, 247212 (2012).
\bibitem{sugano} S. Sugano, Y. Tanabe, and H. Kamimura, {\it Multiplets of Transition-Metal Ions in Crystals} 
(Academic Press, New York, 1970). 
\bibitem{Jackeli} G. Jackeli and G. Khaliullin, Phys. Rev. Lett. \textbf{102}, 017205 (2009).
\bibitem{Jin} H. Jin, H. Jeong, T. Ozaki, and J. Yu, Phys. Rev. B \textbf{80}, 075112 (2009). 
\bibitem{Watanabe} H. Watanabe, T. Shirakawa, and S. Yunoki, Phys. Rev. Lett. \textbf{105}, 216410 (2010).
\bibitem{Martins} C. Martins, M. Aichhorn, L. Vaugier, and S. Biermann, Phys. Rev. Lett. {\bf 107}, 266404 (2011). 
\bibitem{Shirakawa} T. Shirakawa, H. Watanabe, and S. Yunoki, J. Phys. Soc. Jpn. \textbf{80}, SB010 (2011).
\bibitem{Wang} F. Wang and T. Senthil, Phys. Rev. Lett. \textbf{106}, 136402 (2011).
\bibitem{Kanamori} J. Kanamori, Prog. Theor. Phys. \textbf{30}, 275 (1963).

\bibitem{note1}
When a smaller value of $\lambda$ is used, the non-interacting band $m=2$ [i.e., $E_2({\bm k})$] pushes upwards and crosses 
the Fermi energy with hole pockets at around 
${\bm k}=(\pi,\pi)$ and the equivalent momenta (see Fig.~\ref{fig1}). 
However, the Coulomb interactions enhance the effective SOC in ${\tilde H}_0$, which pushes the effective band $m=2$ 
[i.e., ${\tilde E}_2({\bm k})$] downwards away from the Fermi energy. 

\bibitem{note2}
We have checked that our results are not affected semi quantitatively by small change of the one-body parameters 
in $H_0$ as long as only one band crosses the Fermi energy 
after the effect of Coulomb interactions is considered. 

\bibitem{note3} We have checked that in the superconducting region found in this paper the off-diagonal components $\varDelta_{mm'}$ 
($m\neq m'$) 
are negligibly small compared to the diagonal ones. This is easily understood by recalling the non-interacting bands where 
all bands but $m=1$ are fully occupied (see Fig.~\ref{fig1}). 
\bibitem{Sorella} S. Sorella, Phys. Rev. B \textbf{64}, 024512 (2001); M. Casula and S. Sorella, J. Chem. Phys. {\bf 119}, 
6500 (2003); M. Casula, C. Attaccalite, and S. Sorella, J. Chem. Phys. {\bf 121}, 7110 (2004); S. Yunoki and S. Sorella, 
Phys. Rev. B {\bf 74}, 014408 (2006); S. Sorella, M. Casula, and D. Rocca, J. Chem. Phys. {\bf 127}, 014105 (2007). 
\bibitem{note4} We have studied different pairing symmetries, including $s$-wave, extended $s$-wave, $p$-wave, $d_{xy}$-wave, 
and $d_{x^2-y^2}$-wave symmetries, and found that the $d_{x^2-y^2}$-wave singlet is the most stable (most of the optimized 
$\varDelta$'s with other pairing symmetries are found zero) for the parameter region studied. 

\bibitem{note5} 
The explicit form is 
$ 
\varDelta_{mm}=\varDelta_1(\cos k_x-\cos k_y)
+\varDelta_3(\cos 2k_x-\cos 2k_y) \notag 
+2\varDelta_4(\cos 2k_x \cos k_y-\cos k_x \cos 2k_y) 
+\varDelta_5(\cos 3k_x-\cos 3k_y) \label{pairing}
$.

\bibitem{Mesot} J. Mesot, M. R. Norman, H. Ding, M. Randeria, J. C. Campuzano, A. Paramekanti, H. M. Fretwell, A. Kaminski, 
T. Takeuchi, T. Yokoya, T. Sato, T. Takahashi, T. Mochiku, and K. Kadowaki, Phys. Rev. Lett. {\bf 83}, 840 (1999); 
T. Watanabe, H. Yokoyama, K. Shigeta, and M. Ogata, New J. Phys. \textbf{11}, 075011 (2009).

\bibitem{note6} Note that because of the renormalization of the effective SOC in ${\tilde H}_0$ due to the Coulomb interactions, 
the effective band $m$=2 is pushed downward and is located below the Fermi energy for hole doping up to $\sim30$\% when 
$U/t_1$=8. We have checked possible superconducting states including intra- and inter-band pairings and found that none of them 
are stabilized in the hole doped region.


\bibitem{Dagotto} E. Dagotto, A. Nazarenko, and M. Boninsegni, Phys. Rev. Lett. {\bf 73}, 728 (1994); E. Dagotto, A. Nazarenko, 
and A. Moreo, Phys. Rev. Lett. {\bf 74}, 310 (1995). 
\bibitem{deMedici} L. de' Medici, Phys. Rev. B \textbf{83}, 205112 (2011). 
\bibitem{tba} The details of the condensation energy will be discussed elsewhere (H. Watanabe, T. Shirakawa, and S. Yunoki, in preparation).
\bibitem{Vaugier} L. Vaugier, H. Jiang, and S. Biermann, Phys. Rev. B \textbf{86}, 165105 (2012).
\bibitem{Arita} R. Arita, J. Kune\v{s}, A. V. Kozhevnikov, A. G. Eguiluz, and M. Imada, Phys. Rev. Lett. \textbf{108}, 086403 (2012).
\bibitem{Comin} R. Comin, G. Levy, B. Ludbrook, Z.-H. Zhu, C.N. Veenstra, J.A. Rosen, Yogesh Singh, P. Gegenwart, D. Stricker, J.N. Hancock, 
D. van der Marel, I.S. Elfimov, and A. Damascelli, arXiv:1204.4471; B. H. Kim, G. Khaliullin, and B. I. Min, arXiv:1205.3289. 
\bibitem{Giamarchi} T. Giamarchi and C. Lhuillier, Phys. Rev. B \textbf{43}, 12943 (1991); A. Himeda and M. Ogata, Phys. Rev. B \textbf{60}, R9935 (1999).
\bibitem{Kobayashi} K. Kobayashi and H. Yokoyama, Physica C \textbf{470}, 1081 (2010).
\bibitem{Eskes} H. Eskes, G. A. Sawatzky, L. F. Feiner, Physica C {\bf 160}, 424 (1989); 
T. Tohyama and S. Maekawa, J. Phys. Soc. Jpn. {\bf 59}, 1760 (1990). 

\end{thebibliography}
\end{document}